\documentclass[conference]{IEEEtran}

\usepackage{amsmath,amssymb,amsfonts}
\usepackage{bm}
\usepackage{cite}
\usepackage{graphicx}
\usepackage{array}
\usepackage{booktabs}
\usepackage{multirow}
\usepackage{url}
\usepackage{xcolor}
\usepackage{subfig}
\usepackage{url}
\usepackage{hyperref}

\newcommand{\az}{\phi}   
\newcommand{\el}{\theta} 

\begin{document}

\title{Evaluation of gNB Monostatic Sensing  \\ for UAV Use Case}

\author{Steve~Blandino$^{1,2}$, Neeraj Varshney$^{1,2}$, Jian Wang$^3$, Jack Chuang$^3$, Camillo Gentile$^3$, Nada Golmie$^3$
\\
$^{1}$ Associate, National Institute of Standards and Technology (NIST), Gaithersburg, MD. \\
     $^{2}$  Prometheus Computing LLC, Bethesda, MD.\\
     $^{3}$  National Institute of Standards and Technology (NIST), Gaithersburg, MD.
}

\markboth{IEEE Transactions on Wireless Communications (Draft)}%
{Blandino \MakeLowercase{\textit{et al.}}: Standard-Compliant PRS-Based Monostatic Sensing under 3GPP Rel-19 ISAC Channel Model}

\maketitle

\begin{abstract}
3GPP Release 19 has initiated the standardization of integrated sensing and communications (ISAC), 
including a channel model for monostatic sensing, evaluation scenarios, and performance assessment methodologies. 
These common assumptions provide an important basis for ISAC evaluation, but reproducible end-to-end studies still require a 
transparent sensing implementation. 
This paper evaluates 5G New Radio (NR) base station (gNB)-based monostatic sensing for the Unmanned Aerial Vehicle (UAV) 
use case using a 5G NR downlink Cyclic Prefix-Orthogonal Frequency Division Multiplexing (CP-OFDM) 
waveform and positioning reference signals (PRS), 
following 3GPP Urban Macro-Aerial Vehicle (UMa-AV) scenario assumptions. 
We present an end-to-end processing chain for multi-target detection and 3D localization, 
achieving more than 70\,\% detection probability with less than 5\,\% false alarm rate, in the considered scenario. 
For correctly detected targets, localization errors are on the order of a few meters, with a 90th-percentile error of 4\,m and 6\,m 
in the vertical and horizontal directions, respectively. 
To support reproducible baseline studies and further research, we release the  simulator \texttt{5GNRad}, 
which reproduces our evaluation.
\end{abstract}


\section{Introduction}
\label{sec:intro}

The integration of sensing into cellular networks has emerged as a key enabler for 6G, supporting use cases such as aerial surveillance, infrastructure monitoring, transportation safety, and public security \cite{9829746, 11404239}.
Cellular deployments offer widely distributed sensing infrastructure that can complement dedicated radar systems in terms of coverage and cost, while using existing hardware and spectrum resources.

Despite rapid progress in cellular sensing research \cite{11309898, 9921271, 11310754,11205139,10978602,icc25}, performance evaluations remain fragmented. Prior work often relies on simplified propagation assumptions, such as single-bounce targets\cite{11309898, 9921271, 11310754,11205139,10978602,icc25} and negligible clutter\cite{11309898, 9921271,10978602,icc25}, and on idealized hardware models that ignore antenna patterns and residual self-interference\cite{11309898, 9921271, 11310754,11205139,10978602,icc25}. As a result, reproducible and comparable performance evaluations remain limited, and reported results are often difficult to compare objectively across studies.

In parallel with the research community, 3GPP has initiated the standardization of integrated sensing and communication (ISAC) in Technical Specification Group Radio Access Network (TSG RAN) Working Group 1 (WG1), or RAN1, by defining common evaluation scenarios and assumptions. In Release-19, the ISAC channel model extends the baseline 3GPP channel model by introducing sensing targets (STs) characterized by target radar cross section (RCS), along with a dedicated clutter model~\cite{3gppTR38901,11359163, 11396937,11159279}. 
However, the sensing processing chain is outside the scope of 3GPP standardization effort. Thus, while 3GPP specifies the propagation assumptions and evaluation setting used to generate the sensing channel, it does not prescribe how the received signal should be processed into detections and target estimates.

In this work, we evaluate the performance of 5G NR base station (gNB) monostatic sensing for the UAV use case using an existing 5G NR downlink Cyclic Prefix-Orthogonal Frequency Division Multiplexing (CP-OFDM) waveform and the standardized downlink positioning reference signal (PRS) as the sensing signal. The evaluation is grounded in the deployment scenario and assumptions used in 3GPP Release 19 to calibrate the ISAC channel model for UAV sensing targets. 
On top of the standardized channel model, we develop a baseline receiver implementation that includes channel estimation, range-Doppler processing, constant false alarm rate (CFAR) detection, angle estimation, and 3D localization. We also account for practical transceiver impairments by modeling residual self-interference as an additional noise term and by considering realistic antenna patterns and array geometries. 

To enable reproducibility and serve as a common reference point for further research and standards studies, we release the corresponding  baseline simulator, \texttt{5GNRad}\cite{nist_5gnrad}, together with example scenarios and scripts that reproduce the full pipeline and facilitate extensions with new algorithms and assumptions.
Section~\ref{sec:system} presents the sensing system model, including the waveform, array architectures, and the resulting frequency-domain received-signal model.
Section~\ref{sec:channel} then describes the 3GPP Release-19 propagation channel model used to generate the target and background responses.
Section~\ref{sec:processing} details the baseline processing chain that operates on the received signal to produce detections and target parameter estimates.
Section~\ref{sec:simsetup} describes the simulation setup following 3GPP Urban Macro-Aerial Vehicle (UMa-AV) scenario assumptions.
Section~\ref{sec:results} provides the simulation results.
Section~\ref{sec:conclusion} concludes the paper.

\section{5G NR Monostatic Radar}
\label{sec:system}

\begin{figure}[t]
    \centering
    \includegraphics[width=1\linewidth]{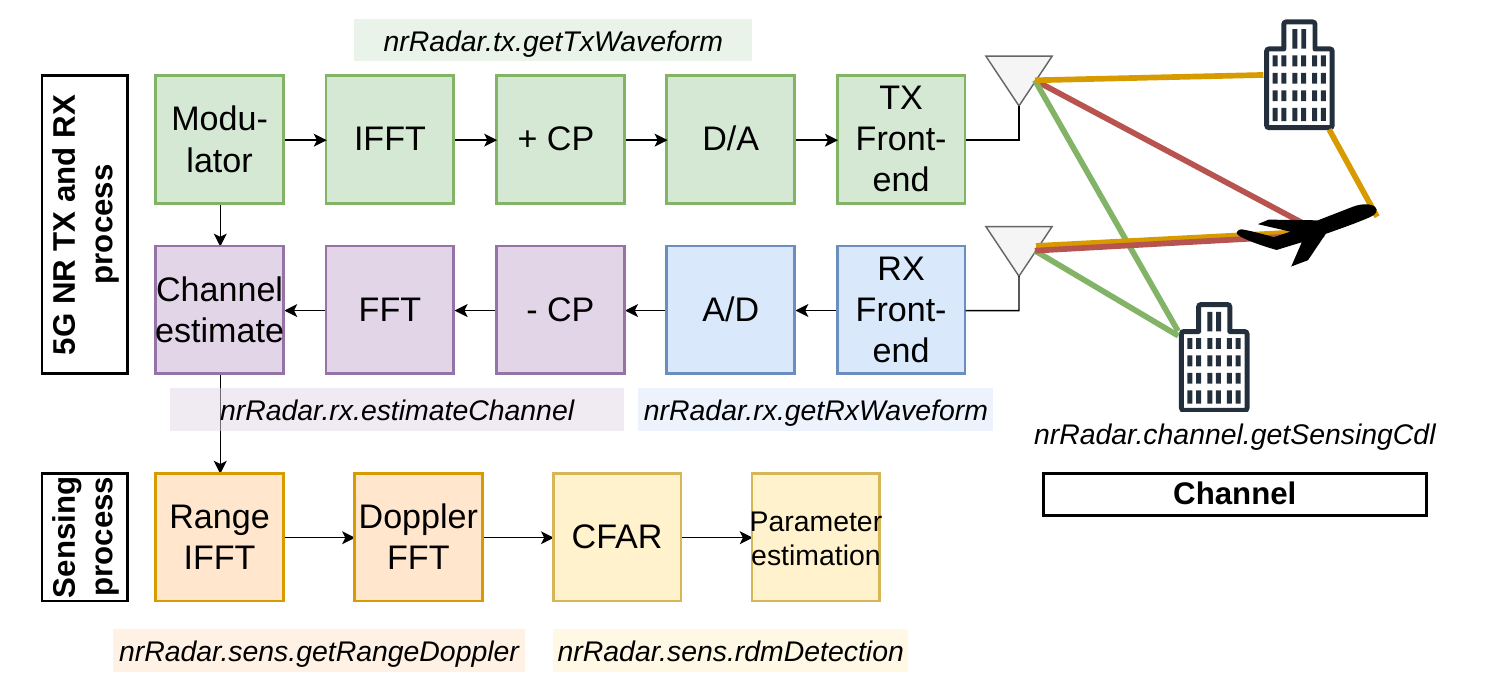}
    \caption{A TRP transmits and receives a CP-OFDM waveform, which propagates through the 3GPP channel model.}
    \label{fig:overview}
    \vspace{-5mm}
\end{figure}

We consider a single gNB operating as a monostatic full-duplex CP-OFDM radar. The transmit--receive point (TRP) denotes the physical radio site of the gNB and comprises co-located transmit and receive  uniform rectangular arrays (URAs) with $N$ antenna elements each. Sensing is performed by transmitting 5G NR positioning reference signals (PRS) and processing the received echoes from sensing targets. The end-to-end signal generation and processing flow is shown in Fig.~\ref{fig:overview}. The processing blocks in Fig.~\ref{fig:overview} are also labeled with the corresponding \texttt{5GNRad}~\cite{nist_5gnrad}  function names.

Let $\Delta f$ denote the subcarrier spacing in Hz and $N_\text{sc}$ the number of subcarriers in the occupied bandwidth. PRS is mapped on a comb of size $K\in\{2,4,6,12\}$, so only every $K$-th subcarrier is active. For PRS OFDM symbol index $\ell$ and subcarrier index $k$, the PRS grid is $s_{k,\ell}=c_{k,\ell}$ if $k\nobreak\hspace{0.08em}\bmod\nobreak\hspace{0.08em}K=\nobreak\hspace{0.08em}0$, and $s_{k,\ell}=0$ otherwise.
 $c_{k,\ell}$ are QPSK-modulated symbols generated from the 31st-order Gold sequence specified in Section~5.2.1 of 3GPP TS~38.211.

PRS occasions are scheduled periodically with period $T_\text{PRS}$, analogous to a pulse repetition interval, and each PRS occasion contains $L_\text{PRS}$ PRS symbols. A coherent processing interval (CPI) consists of $N_\text{CPI}$ PRS occasions spanning time $T_\text{CPI}=N_\text{CPI}T_\text{PRS}$. The baseband OFDM waveform is obtained via Inverse Fast Fourier Transform (IFFT) and cyclic-prefix insertion per symbol.


The TRP transmits a single stream with quasi-omnidirectional coverage over the forward sector. On reception, the array signals are mapped to one or more baseband streams depending on the receive architecture.

For PRS resource element $(k,\ell)$, the transmitted array signal is:
\begin{equation}
\bm x_{k,\ell}=\sqrt{P_\text{TX}}\,\bm f\,s_{k,\ell}\in\mathbb{C}^{N\times 1},
\label{eq:tx_vec}
\end{equation}
where $P_\text{TX}$ denotes the average transmit power per active PRS resource element and $\bm f\in\mathbb{C}^{N\times 1}$ is the fixed transmit weight vector. After CP removal and OFDM demodulation, the $N_\text{RF}$ baseband streams produced by the receive architecture are modeled as:
\begin{align}
\bm y_{k,\ell}
&=\bm W_\text{RF}^{H}\!\left(\bm H_{k,\ell}\bm x_{k,\ell}+\bm n_{k,\ell}+\bm \xi_{k,\ell}\right) \nonumber\\
&=\sqrt{P_\text{TX}}\,\bm W_\text{RF}^{H}\bm H_{k,\ell}\bm f\,s_{k,\ell}
+\bm W_\text{RF}^{H}\bm n_{k,\ell}
+\bm W_\text{RF}^{H}\bm \xi_{k,\ell},
\label{eq:system_model}
\end{align}
where $\bm H_{k,\ell}\in\mathbb{C}^{N\times N}$ is the effective frequency-domain monostatic MIMO channel, as defined in Section \ref{sec:channel}, at subcarrier $k$ and PRS symbol $\ell$, i.e., including antenna patterns and array responses, $\bm n_{k,\ell}\sim\mathcal{CN}(\bm 0,\sigma_n^2\bm I)$ models the receiver thermal noise, and $\bm \xi_{k,\ell}\sim\mathcal{CN}(\bm 0,\sigma_{si}^2\bm I)$ denotes residual self-interference (SI) leakage after cancellation.

The receive architecture is described by an analog combining matrix $\bm W_\text{RF}\in\mathbb{C}^{N\times N_\text{RF}}$ that maps the $N$ element signals into $N_\text{RF}$ baseband streams, as in \eqref{eq:system_model}. Range-Doppler processing is performed per RF chain on the streams $\{y^{(i)}_{k,\ell}\}_{i=1}^{N_\text{RF}}$, producing $N_\text{RF}$ range-Doppler maps.

The receiver architecture determines the number of RF chains $N_\text{RF}$. In the full-digital case, $N_\text{RF}=N$ and $\bm W_\text{RF}=\bm I_N$, so all element signals are available in baseband and range--Doppler processing is performed per element channel. In the hybrid case, $1<N_\text{RF}<N$, the array is partitioned into subarrays, and each RF chain is formed by analog combining within a subarray; range--Doppler processing is then carried out per RF chain and the resulting multi-stream snapshots are used for angle estimation. In the analog case, $N_\text{RF}=1$ and a single analog beam selected from a codebook is applied directly at RF; angle acquisition relies on beam sweeping and the subsequent processing proceeds on a single digital stream.

\section{3GPP Release-19 ISAC Channel Model}
\label{sec:channel}

This section summarizes the 3GPP Release-19 ISAC channel model components used in our evaluation. We focus on monostatic TRP sensing with UAV targets in the UMa-AV scenario. The channel is  described in the time domain through its channel impulse response (CIR), as outlined below. It is then transformed to the frequency domain to obtain the channel matrices  $\bm H_{k,\ell}$  used in the system model in Eq. (\ref{eq:system_model}).

In Release-19, the monostatic channel at the TRP is modeled as the sum of the sensing-target contribution and a monostatic background component. Denoting by $\bm H(\tau,t)$ the time-varying baseband-equivalent MIMO impulse response at delay $\tau$ and time $t$, we write:
\begin{equation}
\bm H(\tau,t) = \bm H^{(\mathrm{ST})}(\tau,t) + \bm H^{(\mathrm{BG})}(\tau,t),
\label{eq:h_time_decomp}
\end{equation}
where $\bm H^{(\mathrm{ST})}(\tau,t)$ aggregates the channels induced by all sensing targets, and $\bm H^{(\mathrm{BG})}(\tau,t)$ models the monostatic background channel, i.e., clutter multipath.

\subsection{Sensing target model and RCS}
\label{sec:rcs_model}
A sensing target is represented by one or more scattering points (SPs). For the small-UAV use case considered here, 3GPP models the target as a single dominant SP with the following monostatic RCS: $\sigma_\mathrm{RCS} = \sigma_M\,\sigma_D\,\sigma_S$,
where $\sigma_M$ is a deterministic mean RCS component, $\sigma_D$ is a deterministic component that can capture angular dependencies, and $\sigma_S$ captures small-scale fluctuations modeled as log-normal in the linear domain \cite{3gppTR38901}.

\subsection{Target channel construction for monostatic sensing}
\label{sec:target_channel}
Each target SP generates multiple rays to capture scattering around the nominal path due to the target structure and its interaction with the environment. The monostatic target channel is obtained by coupling the TRP--SP and SP--TRP propagation segments according to the Release-19 procedure. Large-scale parameters follow the reference deployment scenario, while small-scale parameters are generated for each segment and then combined to form monostatic paths. Weak coupled rays are discarded using a relative-power threshold. The resulting target tapped-delay MIMO impulse response is:
\begin{align}
&\bm H^{(\mathrm{ST})} =
\sum_{q=1}^{Q}\sum_{p=1}^{P_q} \alpha_{q,p}(t)\,
\bm a_\mathrm{rx}\!\left(\Omega^\mathrm{AOA}_{q,p}\right)\,
\bm a_\mathrm{tx}^{\mathsf{H}}\!\left(\Omega^\mathrm{AOD}_{q,p}\right)\,
\delta(\tau-\tau_{q,p}),
\label{eq:target_tdl}
\end{align}
where $Q$ is the number of sensing targets, $P_q$ is the number of retained coupled paths for target $q$, $\tau_{q,p}$, $\Omega^\mathrm{AOA}_{q,p}$ and $\Omega^\mathrm{AOD}_{q,p}$ are respectively the absolute delays and the arrival and departure angles for the $p$-th path for the $q$-th target, and $\bm a_\mathrm{rx}(\cdot)$ and $\bm a_\mathrm{tx}(\cdot)$ denote the receive and transmit array response vectors. The complex gain $\alpha_{q,p}(t)$ includes large-scale attenuation, i.e., path loss, shadow fading, and small-scale fading of both propagation segments, the RCS scaling in Section \ref{sec:rcs_model}, and the Doppler phase due to target motion.

\subsection{Monostatic background channel model}
\label{sec:background_channel}
For monostatic sensing at a co-located TRP, Release-19 introduces a dedicated background channel model intended to represent clutter multipath. The model generates $N_\mathrm{RP}=3$ virtual reference points (RPs) around the TRP with scenario-dependent distance and height distributions. Each TRP--RP link is forced to be Non-Line-of-Sight (NLoS) and is generated using the standard 3GPP clustered-channel procedure for the corresponding reference deployment scenario \cite{3gppTR38901}. 
Because the TRP and virtual RPs are static, the background channel has no geometric time evolution over a CPI. We therefore write $\bm H^{(\mathrm{BG})}(\tau,t)= \bm H^{(\mathrm{BG})}(\tau)$ and omit the explicit time dependence. The overall background impulse response is then formed as:
\begin{equation}
\bm H^{(\mathrm{BG})}(\tau)=\sum_{r=1}^{N_\mathrm{RP}} 10^{-(\mathrm{PL}_r+\mathrm{SF}_r)/20}\,\bm H^{(\mathrm{RP})}_r(\tau),
\label{eq:bg_rp_sum}
\end{equation}
where $\bm H^{(\mathrm{RP})}_r(\tau)$ is the clustered impulse response of the $r$-th TRP--RP link, and $\mathrm{PL}_r$ and $\mathrm{SF}_r$ denote its path loss and shadow fading in dB, respectively.
\section{Baseline Signal Processing Chain}
\label{sec:processing}

This section describes the end-to-end processing chain implemented in \texttt{5GNRad}. The processing chain is not specified by 3GPP, rather, it is a transparent reference implementation used to convert the simulated received PRS observations into detections and parameter estimates. The chain takes the received baseband streams $\bm y_{k,\ell}\in\mathbb{C}^{N_\text{RF}\times 1}$ on PRS resource elements, as defined in \eqref{eq:system_model}, and outputs detections and parameter estimates including range, radial velocity, azimuth/elevation, and 3D position.

\subsection{Channel estimation}
For each active PRS resource element $(k,\ell)$ with $s_{k,\ell}\neq 0$, the receiver forms a least-squares (LS) estimate of the effective frequency-domain channel after transmit precoding and analog combining. Specifically, from \eqref{eq:system_model} we can write
\begin{equation}
\bm y_{k,\ell} = \sqrt{P_\text{TX}}\,\bm g_{k,\ell}\,s_{k,\ell} + \bm \eta_{k,\ell},
\label{eq:y_eff_channel}
\end{equation}
where $\bm g_{k,\ell}\triangleq \bm W_\text{RF}^{H}\bm H_{k,\ell}\bm f\in\mathbb{C}^{N_\text{RF}\times 1}$ is the effective channel vector and $\bm \eta_{k,\ell}\triangleq \bm W_\text{RF}^{H}\bm n_{k,\ell}+\bm W_\text{RF}^{H}\bm \xi_{k,\ell}$ collects thermal noise and residual SI after combining. The LS estimate is then obtained by matched filtering with the known PRS symbol,
\begin{equation}
\widehat{\bm g}_{k,\ell}
= \frac{1}{\sqrt{P_\text{TX}}}\,
\frac{\bm y_{k,\ell}\,s_{k,\ell}^*}{|s_{k,\ell}|^2}
\in\mathbb{C}^{N_\text{RF}\times 1}.
\label{eq:ls_chanest_vec}
\end{equation}
The $i$-th entry $\widehat{g}^{(i)}_{k,\ell}$ corresponds to the channel estimate for RF chain $i$.

\subsection{Per-RF-chain range processing}
Range processing is performed independently on each RF chain $i\in\{1,\ldots,N_\text{RF}\}$. Within each PRS occasion $m$, the PRS symbols are first destaggered in frequency by interleaving the active tones across the $L_\text{PRS}$ PRS symbols into a single frequency vector. We denote the resulting destaggered channel vector by
$\widehat{g}^{(i)}_{k,m}, k\in\mathcal{K}_\text{PRS},$ where $\mathcal{K}_\text{PRS}$ is the set of contiguous frequency indices over the occupied bandwidth after destaggering. In this way, all PRS symbols in the occasion contribute to a single aligned frequency-domain representation.

The complex range profile for RF chain $i$ and PRS occasion $m$ is then computed via an $N_R$-point IFFT across frequency with a tapering window $\bm w_R$:
\begin{equation}
r_i[n,m]
= \frac{1}{\sqrt{N_R}}
\sum_{k\in\mathcal{K}_\text{PRS}}
w_R[k]\,\widehat{g}^{(i)}_{k,m}\,
e^{j2\pi kn/N_R},
\label{eq:range_ifft}
\end{equation}
where $n\in\{0,\ldots,N_R-1\}$ is the delay/range-bin index. The corresponding delay is $\tau[n]=n/(N_R\Delta f)$. The range associated with bin $n$ is $\widehat{R}[n]=\frac{c}{2}\tau[n]$, where $c$ is the speed of light.

\subsection{Static clutter suppression}
To suppress stationary background components, we apply slow-time mean subtraction to the complex range profiles across the CPI:
\begin{equation}
\widetilde{r}_i[n,m]
= r_i[n,m] - \frac{1}{N_\text{CPI}}\sum_{m'=0}^{N_\text{CPI}-1} r_i[n,m'] ,
\label{eq:clutter_mean}
\end{equation}
for each RF chain $i$. This operation attenuates time-invariant components while preserving moving-target returns.

\subsection{Doppler processing and range--Doppler map}
Doppler processing is performed by an $N_D$-point FFT across the PRS-occasion index $m$ at each range bin, using a Doppler window $\bm w_D$:
\begin{equation}
X_i[n,\nu]
= \frac{1}{\sqrt{N_D}}
\sum_{m=0}^{N_\text{CPI}-1}
w_D[m]\,\widetilde{r}_i[n,m]\,
e^{-j2\pi m\nu/N_D},
\label{eq:dopp_fft}
\end{equation}
yielding a complex range--Doppler map per RF chain. Stacking across RF chains forms the range--Doppler snapshot vector
\begin{equation}
\bm X[n,\nu] \triangleq \big[X_1[n,\nu],\ldots,X_{N_\text{RF}}[n,\nu]\big]^\mathsf{T}\in\mathbb{C}^{N_\text{RF}\times 1}.
\label{eq:rd_snapshot}
\end{equation}

The Doppler frequency corresponding to bin $\nu$ and the associated monostatic radial velocity estimate are:
\begin{equation}
\widehat{f}_D[\nu]=\frac{\nu}{N_D T_\text{PRS}}, \qquad 
\widehat{v}_r[\nu]=\frac{\lambda}{2}\,\widehat{f}_D[\nu]=\frac{\lambda\,\nu}{2N_D T_\text{PRS}} .
\label{eq:fd_vr_bin}
\end{equation}

\subsection{CA-CFAR detection}
To perform detection on a single 2D map, the per-chain range--Doppler outputs are projected to a scalar power map through noncoherent integration:
\begin{equation}
P[n,\nu] = \sum_{i=1}^{N_\text{RF}} |X_i[n,\nu]|^2 = \|\bm X[n,\nu]\|_2^2.
\label{eq:proj_rd}
\end{equation}
A two-dimensional cell-averaging CFAR (CA-CFAR) detector is applied to $P[n,\nu]$. For a cell-under-test (CUT) at $(n,\nu)$, training and guard windows are defined, the local noise/clutter power $\widehat{\sigma}^2[n,\nu]$ is estimated as the average over the training ring, and a detection is declared if 
$P[n,\nu] > \gamma\,\widehat{\sigma}^2[n,\nu],$
where $\gamma$ controls the false-alarm–detection tradeoff. A non-maximum suppression step is used to extract a sparse set of peaks and to avoid multiple detections around a dominant lobe.

\subsection{Angle estimation}
\label{sec:doa}
For each detected range--Doppler bin $(n,\nu)$, angle-of-arrival (AoA) estimation uses the complex snapshot vector $\bm X[n,\nu]\in\mathbb{C}^{N_\text{RF}\times 1}$ from \eqref{eq:rd_snapshot}. We consider two angle estimation methods.
\subsubsection{Beamspace FFT}
When the effective receive aperture corresponds to a URA with half-wavelength spacing in both dimensions, an efficient beamspace estimate is obtained via a 2D FFT. The snapshot is reshaped to the appropriate 2D grid $\mathbf{X}_\text{grid}$, windowed by a separable taper $\bm W$, and transformed via a 2-dimensional FFT as $\bm S = \mathrm{FFT2}\{\bm W \odot \bm X_\text{grid}\}$, where $\odot$ denotes term-by-term multiplication. The AoA estimate is thus obtained from the peak of $|\bm S|^2$. 

\subsubsection{Bartlett scan}
For arbitrary spacing where beamspace FFT is not applicable, we use Bartlett beam scanning. Let $\bm a(\az,\el)\in\mathbb{C}^{N_\text{RF}\times 1}$ denote the effective receive steering vector corresponding to angles $(\az,\el)$, which incorporates the analog combining $\bm W_\text{RF}$ and the element pattern and subarray weights. The Bartlett spectrum is:
\begin{equation}
P_B(\az,\el) = \left|\bm a^\mathsf{H}(\az,\el)\,\bm X[n,\nu]\right|^2,
\label{eq:bartlett}
\end{equation}
evaluated over a search grid $(\az,\el)\in\mathcal{G}_\az\times\mathcal{G}_\el$, and the AoA estimate is $(\widehat{\az},\widehat{\el})=\arg\max P_B(\az,\el)$.

\subsection{3D position estimation}
Given the estimated range $\widehat{R}$ and AoA angles $(\widehat{\az},\widehat{\el})$, the unit direction vector is \( \bm u(\widehat{\az},\widehat{\el})  = \left( \cos(\widehat{\theta} )\cos(\widehat{\phi} ), \, \sin(\widehat{\theta} )\cos(\widehat{\phi} ), \, \sin(\widehat{\phi} ) \right)^T \)
For a monostatic TRP located at $\bm p_\text{TRP}\in\mathbb{R}^3$, the 3D target position estimate is
$\widehat{\bm p} = \bm p_\text{TRP} + \widehat{R}\,\bm u(\widehat{\az},\widehat{\el}).
\label{eq:pos_est}$

We enforce sector visibility constraints by discarding estimates outside the sector azimuth bounds or with negative altitude.
\section{3GPP UMa-AV Evaluation Assumptions}
\label{sec:simsetup}

This section summarizes the simulation assumptions used in our evaluation, following the 3GPP ISAC-UAV UMa-AV scenario and the Release-19 ISAC channel model. 

 We consider a single macro site with a single sector whose boresight yaw angle is $\psi=30^\circ$. 
The carrier frequency is $f_c=4$~GHz and the system bandwidth is $B=100$~MHz with subcarrier spacing $\Delta f=30$~kHz. The sensing waveform is NR PRS mapped on a CP-OFDM grid using standardized PRS resource definitions. PRS is configured with period $T_\text{PRS}=1$~ms and $L_\text{PRS}=2$ PRS symbols per occasion. A CPI spans $N_\text{CPI}=128$ PRS occasions, yielding $T_\text{CPI}=128$~ms. The comb size is selected as $K=2$.

The TRP employs an $8\times 8$ URA with $\pm45^\circ$ dual polarization. Two element spacings are considered, namely  $(d_H,d_V)\lambda=(0.5,0.8)\lambda$ and  $(d_H,d_V)\lambda=(0.5,0.5)\lambda$, where $\lambda$ is the wavelength in meters.. The element and array radiation patterns follow 3GPP TR38.901, with a maximum directional element gain of $8$~dBi and $3$-dB beamwidths of $65^\circ$ in both azimuth and elevation. The mechanical tilt points the array toward the horizontal direction. 

Receiver architectures are selected through the number of RF chains $N_\text{RF}$ and the associated analog combining matrix $\bm W_\text{RF}$ as described in Section~\ref{sec:system}. In the full-digital case, $N_\text{RF}=N$ and all element signals are available at baseband. In the hybrid case, the $8\times 8$ URA is partitioned into $4\times 8$ subarrays, each combining two vertically adjacent elements into one RF chain, yielding $N_\text{RF}=32$ baseband streams. This subarraying changes the effective sampling of the aperture, as the vertical spacing between adjacent subarray phase centers becomes $2d_V\lambda$, which equals $\lambda$ when $d_V=0.5$, and equals $1.6\lambda$ when $d_V=0.8$.
Beamforming weights are chosen to maintain quasi-omnidirectional coverage over the forward sector.

Per simulation drop, $Q=5$ small UAV targets are placed within the forward sector coverage area with horizontal positions uniformly distributed over the sector region and altitude uniformly distributed from 25~m to 300~m. The minimum 3D distance between the BS and any target is 10~m, and the minimum 3D separation between targets is 10~m. Target horizontal speed is uniformly distributed from 0~km/h to 180~km/h with random direction in the horizontal plane, while the  vertical speed is 0~km/h. The UAV RCS follows the 3GPP RCS model with $\sigma_M = -12.81$\,dBsm\footnote{dBsm : decibels relative to one square meter, a unit of measurement used in radar to define the Radar Cross Section (RCS)} and $\sigma_D = 1$, while $\sigma_S$ follows a log-normal distribution with $\sigma_{\sigma_S} = 3.74$\,dB.

The TRP-target link is generated using the TRP-aerial UE parameters in TR~36.777  for UMa-AV. The Release-19 ISAC target channel is constructed by concatenating the TRP-target and target-TRP links, followed by path dropping with a $-40$~dB threshold relative to the strongest coupled path. The monostatic background channel is generated using the Release-19 procedure described in Section~\ref{sec:channel}. Residual self-interference is modeled as additional AWGN with configurable power. We consider antenna isolation values of  80~dB with a BS maximum transmit power values of 52~dBm\footnote{dBm: unit of power level expressed using a logarithmic decibel (dB) scale respective to one milliwatt (mW).}.



\section{5G NR Radar Performance Evaluation}
\label{sec:results}
In this section, we present the evaluation results. Unless otherwise stated, results are reported for the baseline configuration with a full-digital antenna array, element spacing $(d_H,d_V)\lambda=(0.5,0.5)\lambda$, and beamspace FFT AOA estimation using a length-128 FFT. The CFAR threshold $\gamma$ is set to 20\,dB and $\sigma_{si} = -\infty$\,dB.

\subsection{Performance Metrics and Estimation Errors}

A detection is reported with estimated parameters (range, radial velocity, angles, and position) and is associated to the closest ground-truth target within the validation gate. We refer to associated detections as true positives ($TP$), unassociated detections as false alarms ($FA$), and missed targets as false negatives ($FN$). The probability of detection is $P_d=TP/(TP+FN)$, i.e., the fraction of true targets that are detected. Following the 3GPP Type-2 definition, let $TP_n$ and $FA_n$ denote the numbers of associated and unassociated detections in drop $n$ under the adopted one-to-one association rule; the false-alarm probability is $P_{FA}= \frac{1}{N_{\rm drop}}\sum_{\substack{n=1\\TP_n+FA_n\neq 0}}^{N_{\rm drop}}\frac{FA_n}{TP_n+FA_n}$, i.e., the fraction of declared detections that are false alarms. We also report the positive predictive value, defined as the fraction of declared detections that are correctly associated, i.e., $P_{TP}=\frac{1}{N_{\rm drop}}\sum_{\substack{n=1\\TP_n+FA_n\neq 0}}^{N_{\rm drop}}\frac{TP_n}{TP_n+FA_n}=1-P_{FA}$, i.e., the fraction of declared detections that correspond to true targets.
We finally report the $F_1$ score, defined as $F_1=2TP/(2TP+FA+FN)$, which summarizes the trade-off between detection probability and false alarms.
Errors are computed over correctly associated detections, i.e., over the $TP$ set. Range and radial-velocity errors are $e_r=|\hat r-r|$ (m) and $e_v=|\hat v_r-v_r|$ (m/s). For position, with ground-truth $\mathbf{p}=[x,y,z]^T$ and estimate $\widehat{\mathbf{p}}$, we use horizontal error $e_H=\|[\hat x,\hat y]^T-[x,y]^T\|_2$ (m) and vertical error $e_V=|\hat z-z|$ (m).

\subsection{Detection and Estimation Performance }
\begin{figure}
  \centering
        \subfloat{\includegraphics[width=0.33\columnwidth]{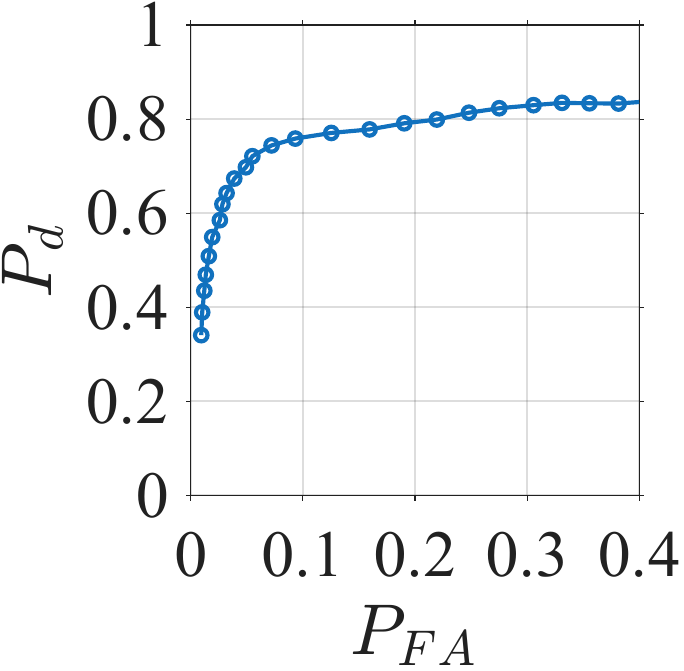}}
        \subfloat{\includegraphics[width=0.32\columnwidth]{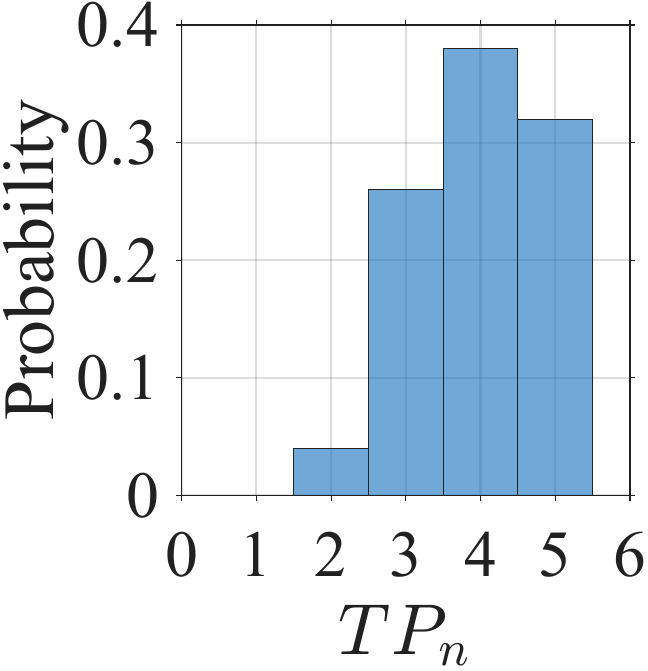}}
        \subfloat{\includegraphics[width=0.32\columnwidth]{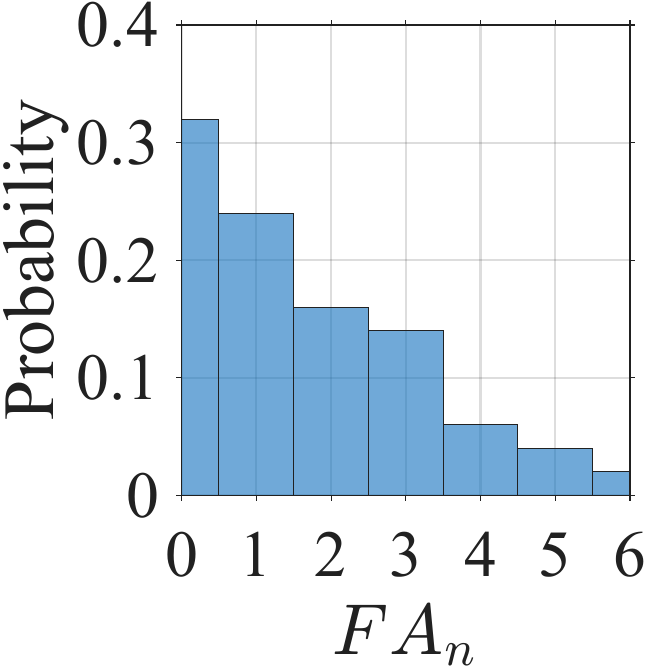}}
    \caption{Detection performance versus CA-CFAR threshold: ROC ($P_d$ vs.\ $P_{FA}$) and distributions of true positives (TP) and false alarms (FP) across drops.
    \vspace{-5mm}
    \label{fig:roc}}
    
\end{figure}
Fig.~\ref{fig:roc} shows the receiver operating characteristic (ROC) obtained by sweeping the CA-CFAR threshold on the range--Doppler map produced by the adopted baseline implementation. At $P_{FA}= 0.05$, the evaluated configuration achieves $P_D= 0.70$. Improving detection performance beyond this point requires a substantial relaxation of the false-alarm constraint: increasing $P_D$ from about $0.70$ to about $0.80$ raises $P_{FA}$ to approximately $0.20$.
To complement the ROC view, Fig.~\ref{fig:roc} also reports the per-drop distributions of $TP_n$ and $FA_n$ at the operating point $P_{FA}= 0.05$. With $Q=5$ targets per drop, the $TP_n$ histogram concentrates around four to five detections per drop, indicating that most targets are detected in typical drops. The $FA_n$ histogram is concentrated at small counts, often zero to a few false alarms per drop.

\begin{figure}
  \centering
        \subfloat{\includegraphics[width=0.32\columnwidth]{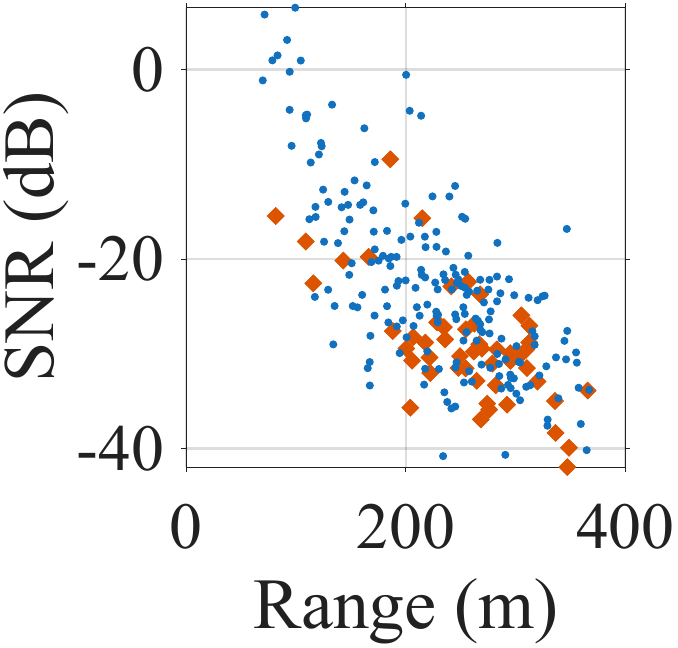}}
        \subfloat{\includegraphics[width=0.315\columnwidth]{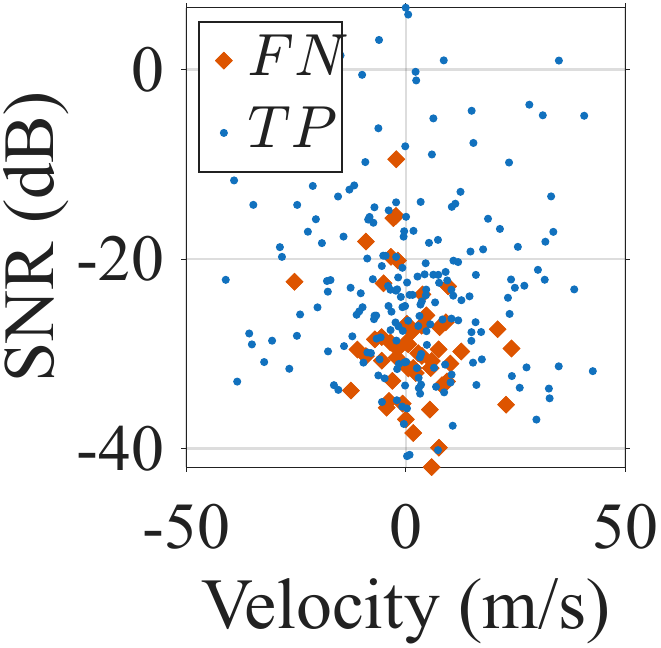}}
        \subfloat{\includegraphics[width=0.32\columnwidth]{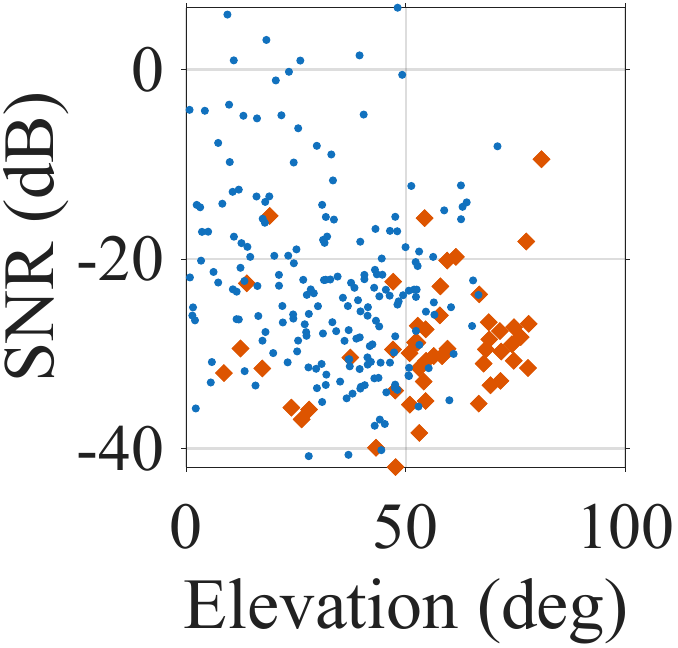}}
    \caption{TP and FN versus range, elevation angle, and radial velocity. \label{fig:miss_scatter}}
    \vspace{-5mm}
\end{figure}

Fig.~\ref{fig:miss_scatter} plots $TP$ and $FN$ events from the simulations versus the corresponding values of SNR, range, elevation, and radial velocity. $FN$s concentrate in the low-SNR regime and become more frequent at larger ranges, consistent with the two-way path loss reducing the target echo power relative to the monostatic background and noise. A second clear trend is observed in elevation: the $FN$ rate increases  for elevation larger than $60^\circ$, which aligns with the effective field of view of the adopted antenna pattern and the reduced array gain near the edge of coverage. Finally, $FN$ events are more prevalent at low radial velocity, suggesting that slow targets are more likely to overlap with residual near-zero-Doppler background components after clutter suppression, thereby reducing peak separability in the range--Doppler domain.

\begin{figure}
  \centering
        \subfloat{\includegraphics[width=0.42\columnwidth]{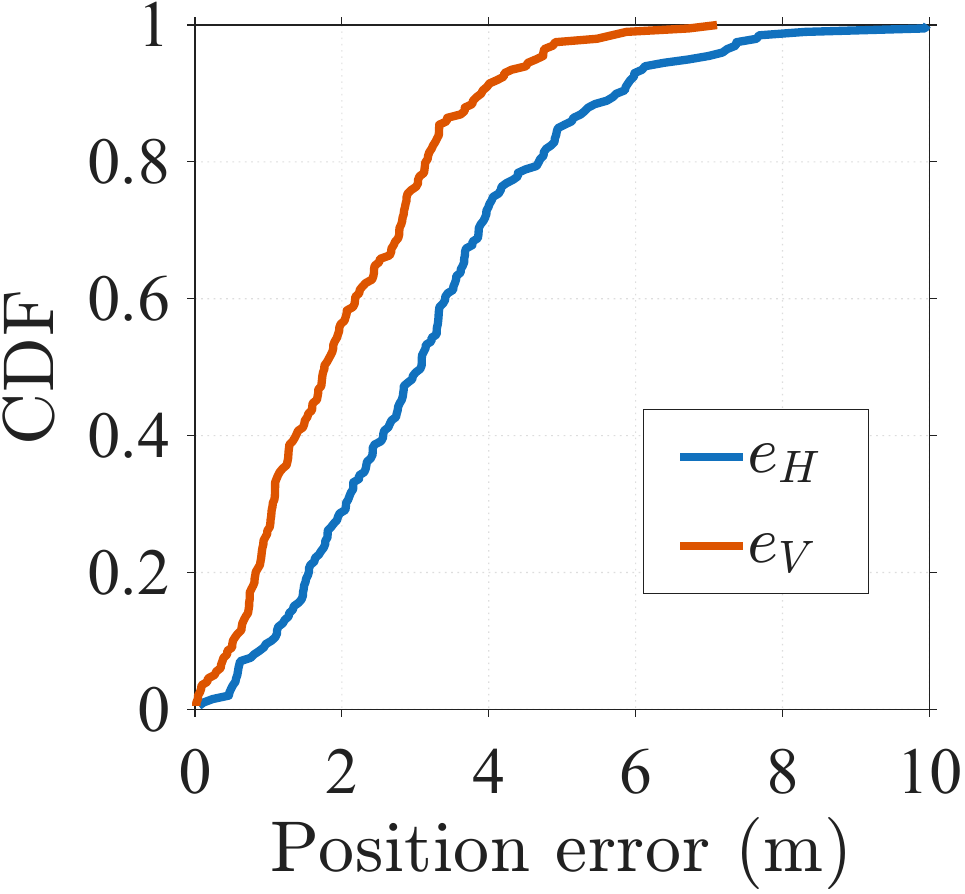}}
        \subfloat{\includegraphics[width=0.42\columnwidth]{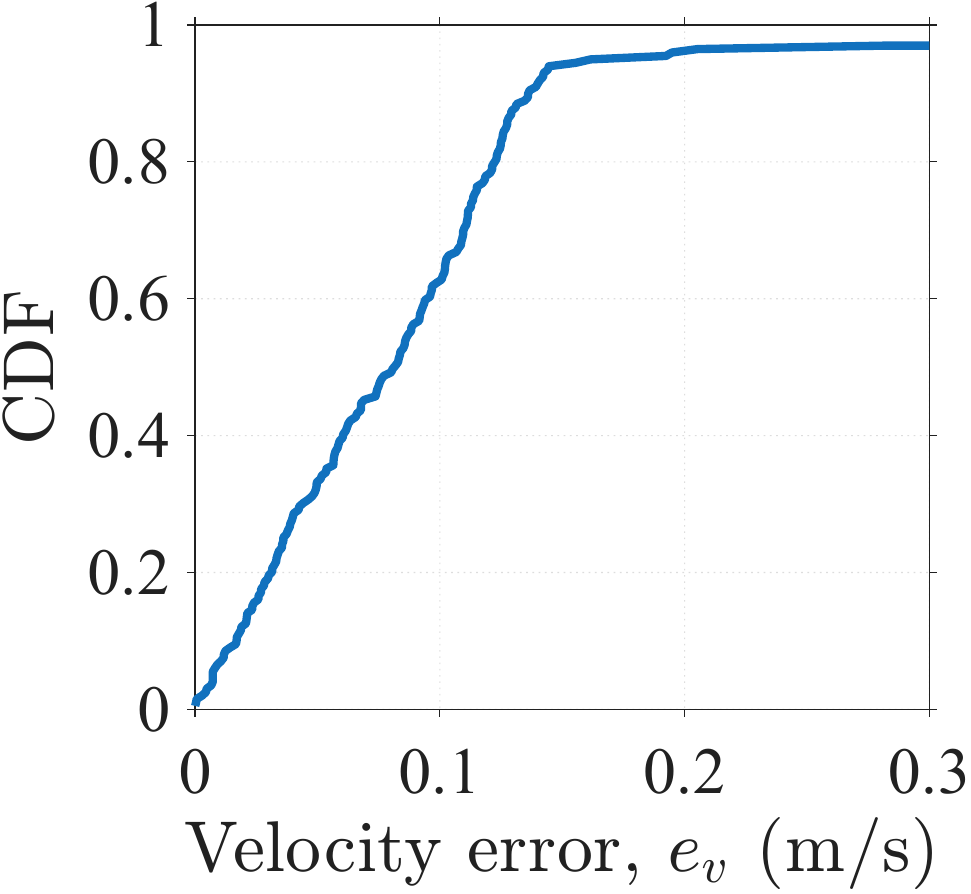}}
            \caption{Empirical CDF of 3D positioning error and velocity error for detected targets.}
                \vspace{-5mm}
    \label{fig:pos_cdf}
\end{figure}

Fig.~\ref{fig:pos_cdf} reports the empirical CDFs of horizontal and vertical position errors together with the radial-velocity error, all computed over correctly associated detections. Under the adopted baseline processing, the evaluated configuration achieves meter-level 3D localization, with 90th-percentile errors of about 6 m horizontally and 4 m vertically, while the radial-velocity error is concentrated at small values, indicating stable Doppler estimation for most correctly detected targets.

\subsection{Sensitivity to residual self-interference and CPI length}
\begin{figure}
  \centering
        \subfloat{\includegraphics[width=0.5\columnwidth]{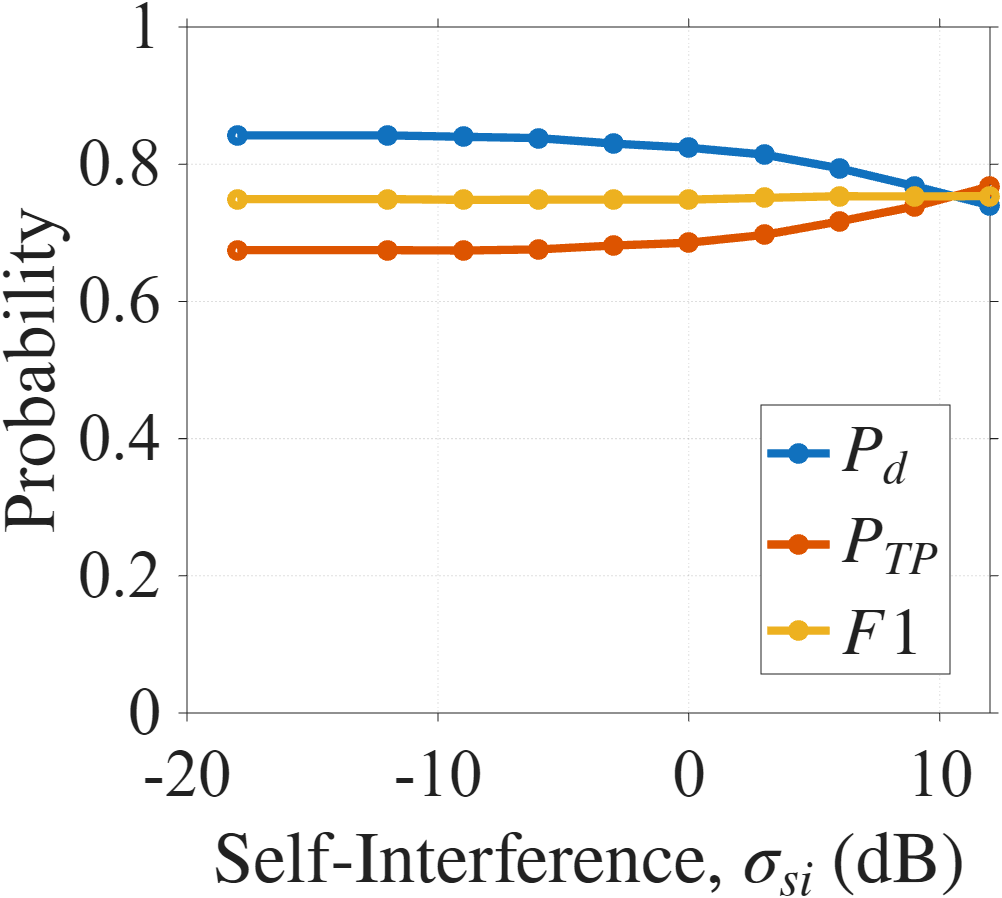}}
        \subfloat{\includegraphics[width=0.50\columnwidth]{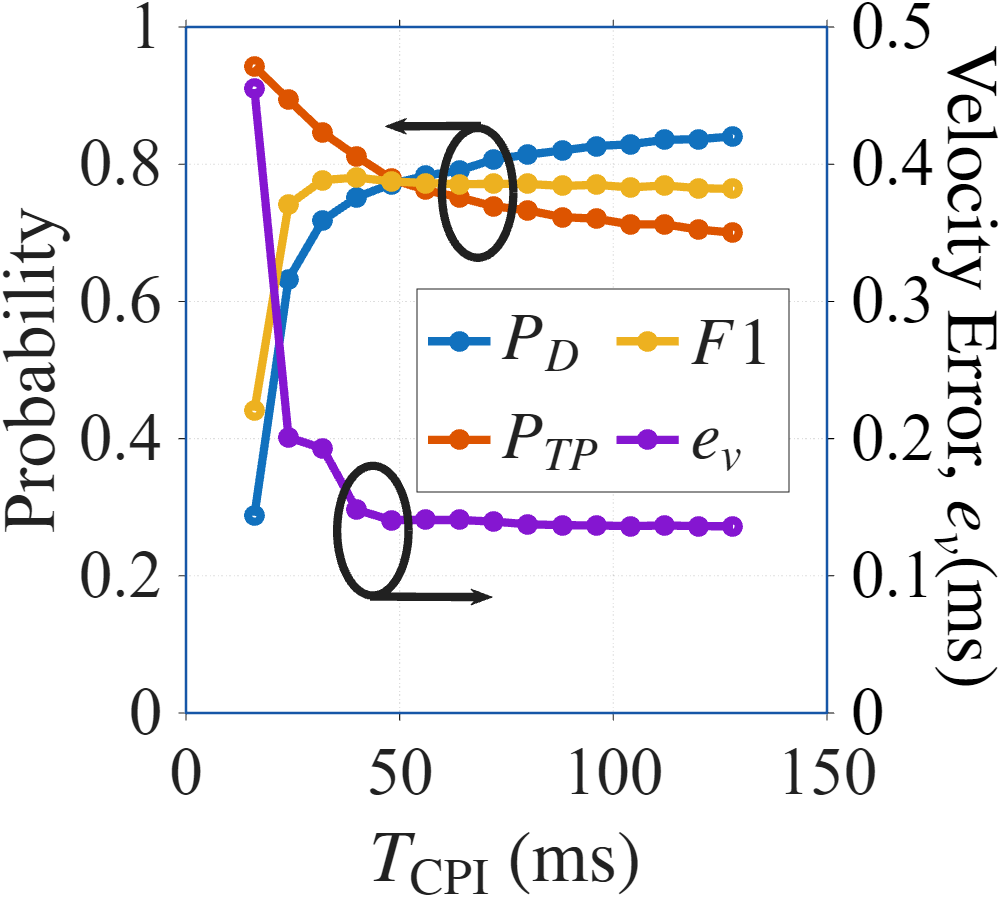}}
    \caption{Left: Sensitivity to residual self-interference. Right: Sensitivity to CPI length.}
        \label{fig:si_cpi}
\vspace{-5mm}
\end{figure}
Fig.~\ref{fig:si_cpi} summarizes the impact of residual SI and CPI length on detection and estimation performance. In Fig.~\ref{fig:si_cpi}(left), increasing the residual SI level reduces $P_d$ because fewer targets exceed the CFAR threshold, while $P_{TP}$ increases as weak noise peaks are covered by the residual SI. The resulting $F_1$ score remains comparatively stable, suggesting that an operating point tuned for a nominal SI condition remains reasonably robust, albeit with an explicit trade-off between $P_D$ and $P_{TP}$. In Fig.~\ref{fig:si_cpi}(right), longer CPI improves $P_d$ and reduces the $90$th-percentile radial-velocity error through increased coherent integration gain and finer Doppler resolution, but slightly degrades $P_{TP}$  since also noise and background channel may experience coherent combination. Overall, $F_1$ and the velocity error saturate beyond $40$~ms, indicating diminishing returns from further CPI extension.

\subsection{Impact of architecture and vertical element spacing}

Table~\ref{tab:det_perf} summarizes the impact of  architecture and vertical element spacing on detection performance. 
We evaluate all configurations using Bartlett AOA estimation with a $1^\circ$ scan step. 
For the full-digital array ($N_{\rm RF}=64$), increasing the vertical spacing from $d_V=0.5$ to $0.8$ reduces both $P_d$ and $P_{TP}$, indicating a modest degradation in separability under wider vertical spacing. Comparing architectures, the hybrid receiver ($N_{\rm RF}=32$) preserves $P_d$ close to the full-digital cases, but exhibits a pronounced drop in $P_{TP}$ and consequently in $F_1$, consistent with an increased false-alarm burden under reduced spatial degrees of freedom. Overall, $d_V$ has a secondary effect relative to the receiver architecture, while hybrid combining primarily impacts false-alarm control rather than detection probability.
\begin{table}[t]
\centering
\caption{Detection performance with $d_H=0.5$. Full-digital: $8\times 8$ URA ($N_{\mathrm{RF}}=64$); hybrid: two vertically adjacent elements per RF chain ($N_{\mathrm{RF}}=32$).}
\label{tab:det_perf}
\footnotesize
\setlength{\tabcolsep}{4pt}
\renewcommand{\arraystretch}{1.1}
\begin{tabular}{@{}ccrrr@{}}
\toprule
$N_\mathrm{RF}$ & $d_V$ & $P_d$ & $P_{TP}$ & $F_1$\\
\midrule
64 & 0.5  & 0.753 & 0.594 & 0.664 \\
64 & 0.8  & 0.722 & 0.569 & 0.637 \\
32 & 0.5  & 0.741 & 0.367 & 0.491 \\
32 & 0.8  & 0.726 & 0.363 & 0.484 \\
\bottomrule
\end{tabular}
\vspace{-1mm}
\end{table}

\section{Conclusion}
\label{sec:conclusion}
This paper evaluated 5G NR gNB-based monostatic sensing for the UAV use case under 3GPP Release-19 assumptions, using standardized channel models, scenarios, and evaluation settings. The results show that 5G NR PRS-based sensing can provide reliable target detection and meter-level localization in the considered UMa-AV scenario.

Beyond the numerical results, this work establishes a reproducible end-to-end evaluation baseline by combining 3GPP standardized procedures with a complete sensing processing chain. Although the sensing receiver is outside the scope of 3GPP standardization, such a reference implementation is important to enable transparent and comparable performance studies. To support this goal, the implementation is released through the open-source \texttt{5GNRad} simulator, together with scripts to reproduce the reported results.
\vspace{-1mm}

\bibliographystyle{ieeetr}
\bibliography{refs.bib}

@INPROCEEDINGS{11309898,
  author={Sagduyu, Yalin E. and Davaslioglu, Kemal and Erpek, Tugba and Kompella, Sastry and Anderson, Gustave and Ashdown, Jonathan},
  booktitle={MILCOM 2025 - 2025 IEEE Military Communications Conference (MILCOM)}, 
  title={{MULTI-SCOUT}: Multistatic Integrated Sensing and Communications in {5G} and Beyond for Moving Target Detection, Positioning, and Tracking}, 
  year={2025},
  volume={},
  number={},
  pages={693-698},
  doi={10.1109/MILCOM64451.2025.11309898}}

@INPROCEEDINGS{11310754,
  author={Blandino, Steve and Golmie, Nada and Sahoo, Anirudha and Nguyen, Thao and Ropitault, Tanguy and Griffith, David and Sonny, Amala},
  booktitle={MILCOM 2025 - 2025 IEEE Military Communications Conference (MILCOM)}, 
  title={{Detecting Airborne Objects with 5G NR Radars}}, 
  year={2025},
  volume={},
  number={},
  pages={1260-1265},
  doi={10.1109/MILCOM64451.2025.11310754}}

@INPROCEEDINGS{11205139,
  author={Wypich, Marek and Zielinski, Tomasz P.},
  booktitle={2025 IEEE Radar Conference (RadarConf25)}, 
  title={{OFDM}-Based Passive Radar Using Reference Signals and User Data}, 
  year={2025},
  volume={},
  number={},
  pages={847-852},
  doi={10.1109/RadarConf2559087.2025.11205139}}

@ARTICLE{11359163,
  author={Hong, Wei and Zhang, Zhenyu and Xi, Wei and Liu, Yunfeng and Li, Yingyang and Huang, Qiheng and Zhao, Qun and Zhou, Juejia and Zhang, Jianhua and Li, Yong},
  journal={IEEE Journal on Selected Areas in Communications}, 
  title={Integrated Sensing and Communication ({ISAC}) Channel Model Toward {3GPP} {6G} Standardization: Modeling, Validation, and Application}, 
  year={2026},
  volume={44},
  number={},
  pages={3459-3472},
  doi={10.1109/JSAC.2026.3656257}}

@software{nist_5gnrad,
  author  = {{National Institute of Standards and Technology (NIST) Communications Technology Laboratory}},
  title   = {5{G} {N}ew {R}adio {R}adar ({5GNR}ad)},
  note   = {Computer software. Available at: \url{https://github.com/usnistgov/5GNRad}}
}

@ARTICLE{11159279,
  author={Zhang, Yuxiang and Zhang, Jianhua and Gong, Huiwen and Hu, Xidong and Zhang, Jiwei and Xing, Hongbo and Luo, Shilin and Xiong, Yifeng and Yu, Li and Yuan, Zhiqiang and Liu, Guangyi and Jiang, Tao},
  journal={IEEE Journal on Selected Areas in Communications}, 
  title={A Unified {RCS} Modeling of Typical Targets for {3GPP} {ISAC} Channel Standardization and Experimental Analysis}, 
  year={2026},
  volume={44},
  number={},
  pages={702-716},
  doi={10.1109/JSAC.2025.3608732}}

@ARTICLE{11396937,
  author={Yang, Wenfei and Cardona, Narcis and Li, Jian and Yu, Ziming and Xia, Jinhuan and Zhang, Zhening and Shao, Jiafeng and Zhu, Peiying},
  journal={IEEE Communications Standards Magazine}, 
  title={Channel Modeling Toward {6G} in {3GPP} Release 19}, 
  year={2026},
  volume={},
  number={},
  doi={10.1109/MCOMSTD.2026.3660670}}

@ARTICLE{11404239,
  author={Qaisar, Muhammad Umar Farooq and Yuan, Weijie and Günlü, Onur and Riihonen, Taneli and Cui, Yuanhao and Zhang, Lin and Gonzalez-Prelcic, Nuria and Renzo, Marco Di and Han, Zhu},
  journal={IEEE Transactions on Network Science and Engineering}, 
  title={The Role of {ISAC} in {6G} Networks: Enabling Next-Generation Wireless Systems}, 
  year={2026},
  volume={},
  number={},
  doi={10.1109/TNSE.2026.3666665}}

@ARTICLE{9829746,
  author={Wang, Jian and Varshney, Neeraj and Gentile, Camillo and Blandino, Steve and Chuang, Jack and Golmie, Nada},
  journal={IEEE Internet of Things Journal}, 
  title={{Integrated Sensing and Communication: Enabling Techniques, Applications, Tools and Data Sets, Standardization, and Future Directions}}, 
  year={2022},
  volume={9},
  number={23},
  doi={10.1109/JIOT.2022.3190845}}

@ARTICLE{9921271,
  author={Wei, Zhiqing and Wang, Yuan and Ma, Liang and Yang, Shaoshi and Feng, Zhiyong and Pan, Chengkang and Zhang, Qixun and Wang, Yajuan and Wu, Huici and Zhang, Ping},
  journal={IEEE Transactions on Vehicular Technology}, 
  title={{5{G} {PRS}-Based Sensing: A Sensing Reference Signal Approach for Joint Sensing and Communication System}}, 
  year={2023},
  volume={72},
  number={3},
  doi={10.1109/TVT.2022.3215159}}

@techreport{3gppTR38901,
  author       = {{3GPP}},
  title        = {{Study on channel model for frequencies from 0.5 to 100 GHz (Release 19.2)}},
  institution  = {{3rd Generation Partnership Project (3GPP)}},
  number       = {{TR 38.901 V16.1.0}},
  year         = {2026},
  month        = {January}
}

@INPROCEEDINGS{10978602,
  author={Khosroshahi, Keivan and Sehier, Philippe and Mekki, Sami and Suppa, Michael},
  booktitle={2025 IEEE Wireless Communications and Networking Conference (WCNC)}, 
  title={Localization Accuracy Improvement in Multistatic {ISAC} with {LoS/NLoS} Condition Using {5G} {NR} Signals}, 
  year={2025},
  volume={},
  number={},
  pages={1-6},
  doi={10.1109/WCNC61545.2025.10978602}}

@INPROCEEDINGS{icc25,
  author={Ropitault, Tanguy and Blandino, Steve and Griffith, David and Sahoo, Anirudha and Nguyen, Thao and Golmie, Nada},
  booktitle={2025 IEEE International Conference on Communications Workshops (ICC Workshops)}, 
  title={{Characterization of Monostatic Base Stations
Sensing Resolution Using 5G Reference Signals}}, 
  year={2025},
  volume={},
  number={},
  pages={},
  doi={}}

\end{document}